\documentclass[lettersize,journal]{IEEEtran}

\usepackage{amsmath,amsfonts}
\usepackage{array}
\usepackage{textcomp}
\usepackage{stfloats}
\usepackage{url}
\usepackage{graphicx}
\usepackage{MnSymbol}
\usepackage{color}
\usepackage{comment}
\usepackage{acronym}
\usepackage{subcaption}
\usepackage{empheq}
\usepackage[font=small]{caption}
\usepackage{enumitem}
\usepackage{cite}

\newcommand{\bs}[1]{\boldsymbol{#1}}
\newcommand{\wqss}{\bs{\omega}_{\rm \scriptscriptstyle QSS}}
\newcommand{\wu}{\bs{\omega_\upsilon}}
\newcommand{\flux}{\bs{\varphi}}
\newcommand{\voltage}{\bs{\upsilon}}
\newcommand{\vorticity}{\mathbf{w}}
\newcommand{\Clarke}{$\alpha\beta\gamma$}
\newcommand{\uvec}[1]{\bs{e}_{#1}}
\newcommand{\wo}{\omega_o}

\acrodef{pll}[PLL]{Phase-Locked Loop}
\acrodef{qss}[QSS]{Quasi Steady-State}
\acrodef{dfig}[DFIG]{Doubly-Fed Induction Generator}
\acrodef{dft}[DFT]{Discrete Fourier Transform}
\acrodef{pmu}[PMU]{Phasor Measurement Unit}
\acrodef{rl}[RL]{Reinforcement Learning}

\begin{document}

\title{Quasi Steady-State Frequency}

\author{Joan Guti{\'e}rrez-Florensa,~\IEEEmembership{Student,~IEEE,}  {\'A}lvaro Ortega,~\IEEEmembership{Member,~IEEE,} Lukas Sigrist,~\IEEEmembership{Member,~IEEE,} and  Federico Milano,~\IEEEmembership{Fellow,~IEEE}
  \thanks{J.~Guti{\'e}rrez-Florensa and F.~Milano are with the School of Elec.~\& Electron.~Eng., University College Dublin, Dublin, D04V1W8, Ireland.  e-mails: joan.gutierrezflorensa1@ucdconnect.ie, federico.milano@ucd.ie, }%
  \thanks{{\'A}.~Ortega and L.~Sigrist are with School of Engineering, Comillas Pontifical University, 28015, Madrid, Spain. e-mails: \{aortega, lsigrist\}@comillas.edu}%
\thanks{This work is supported by the Science Foundation Ireland (SFI) by funding J.~Guti{\'e}rrez-Florensa under NexSys project, Grant No.~21/SPP/3756.}
\vspace{-7mm}
}

\maketitle

\begin{abstract}
  Accurate frequency estimation is critical for the control, monitoring and protection of electrical power systems.  This paper introduces the novel concept of \textit{\ac{qss} frequency} as a quantity that fills the gap between stationary and instantaneous frequency.  \ac{qss} frequency coincides with the fundamental frequency of an AC voltage in any stationary conditions, including unbalanced and non-sinusoidal, and is able to capture the time-varying fundamental frequency in transient conditions.  The paper also proposes a metric borrowed from fluid dynamics, namely, the time derivative of the circulation, to define the scope of validity of the \ac{qss} frequency.  Analytical examples as well as a case study based on a fully-fledged EMT model of the IEEE 39-bus system serve to illustrate, respectively, the properties of the \ac{qss} frequency and its behavior in transient conditions.
\end{abstract}


\section{Introduction}
\label{sec:intro}

\subsection{Motivation}

The meaning of frequency in transient conditions have been widely discussed in the literature, and how to properly define it is still an open question.  The core of the problem is that, on the one hand, frequency is well defined only for a periodic stationary signal (let us call it \textit{Fourier frequency}).  On the other hand, in electrical power systems applications, such as active power control, the estimation of frequency is required during a transient for which the measured signal, typically the voltage at a relevant bus, is not stationary.  The quantity that is usually estimated in transient conditions is known as \textit{instantaneous frequency} \cite{IEEE60255}.  Fourier and instantaneous frequencies coincide only in stationary conditions.  During transients and other common operating conditions such as stationary unbalanced and non-sinusoidal, these two quantities not only do not coincide but give rise to inconsistencies so hard to explain that have been described as ``paradoxes'' \cite{cohen1995time}.  In turn, Fourier and instantaneous frequencies may be treated as two different quantities, and the one that most closely aligns with the assessed phenomenon depends on the specific nature of the application \cite{Mandel}.  This paper proposes a novel quantity, namely, the \textit{\ac{qss} frequency}, that fills the gap between Fourier and instantaneous frequencies.

\subsection{Literature Review}

There exists several approaches to estimate frequency during electrical transients. 
 These include, among others, \acp{pll}, 
\cite{liu2014three, Zhong2023}, \ac{dft}
, \cite{agrez2002weightedDFT, romano2013interpolateddft, song2022fast}, Kalman filters, \cite{reza2016accurate, nie2019detection}, least squares, \cite{giarnetti2015non, pradhan2005freq}, and adaptive notch filters, \cite{wilches2020method, mojiri2007estimation}.
\acp{pll} are a common solution for grid synchronization and frequency control, whereas Fourier-based methods are generally employed in monitoring and state-estimation.  The working principle of \acp{pll} is to estimate the instantaneous frequency through the definition of two voltage components, typically in the $dq$-axis reference frame.  Fourier-based methods use a mobile window assuming a periodic signal.  Both approaches have issues.

\begin{figure}[htb]
  \centering
  \resizebox{0.9\linewidth}{!}{\includegraphics{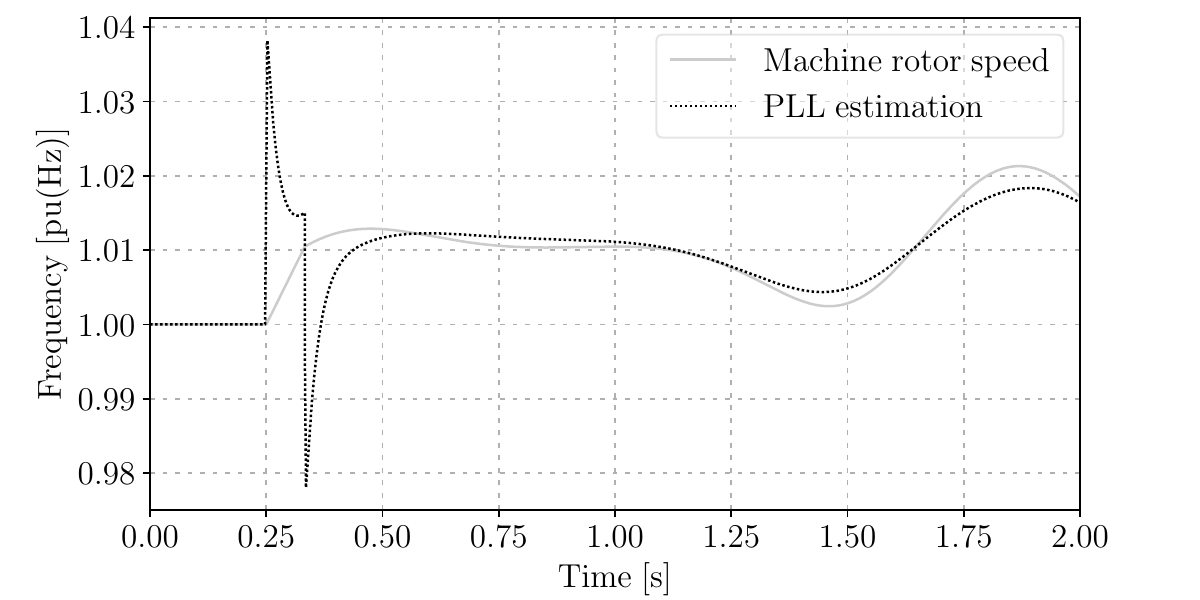}}
  \caption{Difference between rotor speed and the measured electrical frequency at the terminal bus of synchronous machine in a transient following a short circuit and its clearance close to the machine.  Note that the frequency at the terminal bus is not perfectly equal to the machine angular speed in transient condition due to the internal impedance of the machine (see \cite{freqdiv} for a theoretical appraisal and \cite{8586353} for an experimental validation).  Moreover the terminal bus frequency shows fast transients that are not visible in the mechanical angular speed as they are filtered  by the rotor inertia. The comparison only aims at showing the inconsistencies of the spikes of the frequency measured by the \ac{pll} following the fault occurrence and clearance.}
  \label{fig:wscc}
  \vspace{-2mm}
\end{figure}

An issue of the estimation of the instantaneous frequency by means of \acp{pll} is intrinsic to any control loop, that is, the better the noise is filtered, the higher the delay with which the signal is estimated.  A tradeoff is thus needed.  Moreover, the estimation based on \acp{pll} is biased by the well-known issue of showing spikes in correspondence of abrupt events occurring in the system.  Figure \ref{fig:wscc} illustrates this situation by comparing the electrical frequency estimated with a \ac{pll} at the terminal bus of a synchronous machine with the rotor speed of the machine itself.  The spikes appear in the \ac{pll} estimation after a fault and after the line trip that clears the fault.  However, the rotor speed is smooth and does not show spikes.  It is thus legitimate to wonder whether the spikes represent a physical variation of the electrical frequency or a mere limitation of the \ac{pll}.  This is also still an open question.

Fourier-based approaches are generally based on the fast discrete version of the Fourier transform.  These approaches work because, even if the fundamental frequency is not perfectly constant, it never varies too much in power systems, and a window of 20 ms for systems at 50 Hz (16.7 ms for systems at 60 Hz) provides a reasonable representation of the signal.  However, the window creates several problems, such as aliasing and spectral leakage, which, ultimately, cannot be completely solved.  

The coexistence of several approaches to estimate the frequency indicates that no approach is a clear winner.  In \cite{measurements_low_inertia}, various conventional and experimental frequency measurement methods are tested for a variety of robust-demanding situations according to the standard in \cite{IEEE60255} and the tests proposed in \cite{Rietveld}.  The conclusion is that no approach is able to satisfy all the tests but each one can be suggested for different use cases.  Most of these methods use filters or correctors to reduce the measurement oscillations and, in some cases, they return different results for the same case.  This suggests that more research has to be done.  In particular, we believe that a key aspect of most unresolved issues are the conceptual differences between Fourier and instantaneous frequency.

In \cite{paradoxes}, the authors try to conciliate the inconsistencies and paradoxes of Fourier and instantaneous frequencies through a geometrical approach, which identifies the instantaneous frequency as the curvature of an abstract trajectory, where the measured voltage take the meaning of a ``velocity'' \cite{milano2021geometrical}.  The main conclusion of this interpretation is that the instantaneous frequency is a \textit{geometric invariant}, that is, a quantity the value of which does not change no matter the coordinates (measurements) that are utilized \cite{milano2021applications}.  Being an invariant, the instantaneous frequency is always ``right.''  On the other hand, the Fourier frequency is a \textit{coordinate} without any special geometrical meaning except for the case of a stationary sinusoidal balanced voltage.  This conclusion is not fully satisfactory, as the instantaneous frequency is not always the quantity that one expects to obtain, specifically, in non-sinusoidal or imbalanced stationary signals \cite{milano2025affine}.

In summary, the Fourier frequency can in principle be defined only in ideal stationary conditions, but intuitively represents what one would expect to be the fundamental frequency of a signal that is ``almost'' stationary after removing all harmonics and unbalances.  On the other hand, the instantaneous frequency is always valid and its meaning can be rigorously shown to be a precise geometric quantity but often lacks an intuitive meaning and can be deceiving in simple situations such as, for example, a stationary voltage with a small negative sequence component.

Recently, the last author of this work proposed a breakdown of the geometrical frequency into terms that are commonly utilized in fluid dynamics, namely, a vorticity, a time-dependent term, and a distortion (strain) term \cite{Lagrange_FM}.  The vorticity is proportional to the part of the velocity of a fluid line that represents a rigid-body rotation.  In this paper, we show that the vorticity is the term that contains the intuitive information of fundamental frequency of a non-stationary signal.

\vspace{-2mm}
\subsection{Contributions}

This paper poses the theoretical foundation for the definition of a novel quantity, namely the \acf{qss} frequency, which aims at filling the conceptual gap between the Fourier and the instantaneous frequencies.  The theoretical foundation of the definition of \ac{qss} frequency is based on a theorem by Cauchy on vorticity \cite{Cauchy} as well as on the ability provided by differential geometry to estimate the time-varying period of a measured voltage. The \ac{qss} is valid in transients, but retains the intuitive meaning of a ``slowly'' changing fundamental frequency, and it is constant (in fact, equal to the fundamental frequency) for a stationary voltage, even if it is not balanced or sinusoidal.    

Additionally, the \ac{qss} frequency comes with an embedded metric, formally the time derivative of the circulation, which we duly define in the remainder of the paper.  This metric allows establishing the threshold of the validity of the \ac{qss} frequency as an estimation of the instantaneous fundamental frequency.   We show that the \ac{qss} frequency is mathematically defined only if the time derivative of the circulation is below a given threshold.  The proposed metric also clarifies whether the spikes observed for the instantaneous frequency (see Fig.~\ref{fig:wscc}) are physical or just a consequence of the limitations of the \ac{pll} control loop and allows defining in a precise analytical way which parts of the trajectory of the \ac{qss} frequency during a power system transient retain the meaning of a slowly-varying fundamental frequency.

\vspace{-3mm}

\subsection{Paper Organization}

This paper is organized as follows. Section \ref{sec:outlines} recalls relevant concepts from geometric frequency and frequency decomposition from Lagrange derivative equivalence.  This section provides the background for the subsequent developments in this work.  Section \ref{sec:cauchy} includes the formulation and formal definition of \ac{qss} frequency.  Section \ref{sec:circulation} introduces the time derivative of circulation and the proposed use as a metric to determine the existence of a fundamental frequency.   Section \ref{sec:examples} presents a variety of analytical and numerical examples that illustrate the proposed \ac{qss} frequency and metric.  Section \ref{sec:casestudies} examines the behavior of the proposed quantities in a set of case studies.  Section \ref{sec:remarks} provides concluding remarks.  Finally, conclusions and future work are given in Section \ref{sec:conclusions}.

\vspace{-3mm}

\subsection{Notation}

Scalar quantities are represented by italic font, e.g., $a$ or $A$, vectors by bold lower case, e.g., $\bs{a}$, tensors and matrices by bold upper case, e.g., $\bs{A}$.  The transpose of a vector $\bs{a}$ is denoted with $\bs{a}^T$ and the time derivative of a quantity $a$ is indicated with $a'$.  Unless indicated otherwise, quantities are assumed to be time dependent.


\section{Background}
\label{sec:outlines}

This section recalls the definitions of relevant quantities that are utilized in developments of the paper.  These key concepts include \textit{geometric frequency} and \textit{vorticity} as a component of the geometric frequency.

Consider the magnetic flux, $\flux$, as a three-dimensional space curve. Its dynamics are described by:
\begin{equation}
  \flux'=\voltage \, ,
  \label{eq:flux}
\end{equation}
where $\voltage$ is the voltage vector.

This interpretation allows us using the framework of differential geometry to define relevant invariant quantities.  In the context of this paper, the arc length, $s$, and the curvature, $\kappa$, are particularly relevant and are defined as follows \cite{Stoker}:
\begin{equation}
  s = \int_0^t \sqrt{\flux'\cdot\flux'} \,d\tau =
  \int_0^t |\voltage| \, d\tau \, ,
  \label{eq:s}
\end{equation}
\begin{equation}
  \kappa=\frac{|\flux'\times \flux''|}{|\flux'|^{3}}=
  \frac{|\voltage\times \voltage'|}{|\voltage|^{3}} \,,
  \label{eq:k}
\end{equation}
where $\cdot$ and $\times$ are used to represent the inner and cross products of vectors, respectively, as follows:
\begin{equation}
    \label{eq:cross}
   \bs x \cdot \bs y = \sum_{i=1}^3 x_i \, y_i \, , \qquad \bs x \times \bs y = 
   \left |
   \begin{matrix}
     \bs e_1 & \bs e_2 & \bs e_3 \\
     x_1 & x_2 & x_3 \\
     y_1 & y_2 & y_3
   \end{matrix}
   \right | ,
\end{equation}
with $\bs x = x_1 \bs e_1 + x_2 \bs e_2 + x_3 \bs e_3$ and $\bs y = y_1 \bs e_1 + y_2 \bs e_2 + y_3 \bs e_3$, and where $(\bs e_1, \bs e_2,  \bs e_3)$ is an orthonormal basis.

The geometric framework above is used in \cite{milano2021geometrical} to define the concept of geometric frequency as an invariant quantity defined as a multivector composed by a translation, $\rho_\upsilon$, and a rotation, $\wu$.  The latter term is defined as:
\begin{equation}
  \wu = \frac{\voltage\times \voltage'}{|\voltage|^2} \, ,
  \label{eq:wu}
\end{equation}
where $|\wu| = \kappa \, |\voltage|$ is an estimation of the instantaneous frequency of the voltage \cite{milano2021geometrical}.  Equation \eqref{eq:wu} indicates that $\wu$ can be calculated exclusively based on voltage measurements.  

Let us now assume that $\wu = \wu(t, \flux)$.  Then, $\wu$ can be shown to be composed of various terms with precise  physical meanings borrowed from fluid dynamics, as follows \cite{Lagrange_FM}:
\begin{align}
    \nonumber
    \wu &= \frac{\voltage \times \partial_t\voltage}{|\voltage|^{2}} +
    \frac{\voltage \times (\bs{R}\voltage)}{|\voltage|^{2}}-\frac{1}{2}\frac{((\nabla\times\voltage)\cdot\voltage)\voltage}{|\voltage|^2} + 
    \frac{1}{2}(\nabla\times\voltage) \\   \label{eq:lagrange}
    &= \bs{\omega}_t+\bs{\omega}_r+\bs{\omega}_r-\frac{1}{2}\bs{\omega}_\tau\frac{\voltage}{|\voltage|}+\frac{1}{2}\vorticity \, ,
\end{align}
where $\partial_t$ indicates the partial derivative with respect to time; $\bs{\omega}_t$ is related to the non-stationary conditions of the voltage; $\bs{\omega}_r$ is related to the voltage ``distortion,'' characterized by matrix $\bs{R}$; $\bs{\omega}_\tau$ is the torsional frequency; and $\vorticity$ is the vorticity, given by the curl ($\nabla \times$) of the voltage and represents the rotation of a rigid body.  The details of the steps required to obtain the terms that appear in \eqref{eq:lagrange} are given in Appendix \ref{app:lagrange}.

The relevance of this decomposition, in the context of this paper, lies in its ability to theoretically isolate the term associated with rigid-body rotation and what we identify as the inherent frequency of the signal. To better illustrate this point, we provide below an interpretation of \eqref{eq:lagrange} through  representative cases.

\begin{figure}[htb]
  \centering
  \subfloat[]{\includegraphics[scale=0.49]{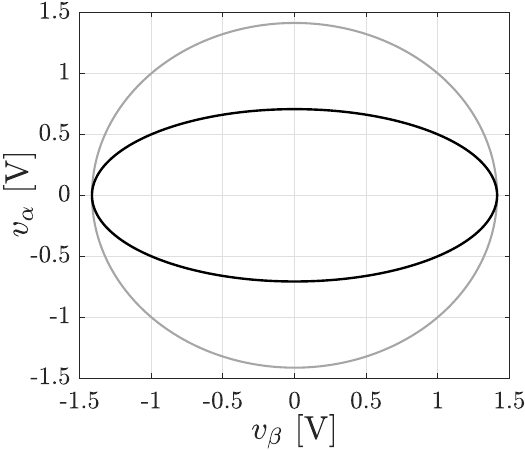}\label{fig:ellipse}}
  \subfloat[]{\includegraphics[scale=0.49]{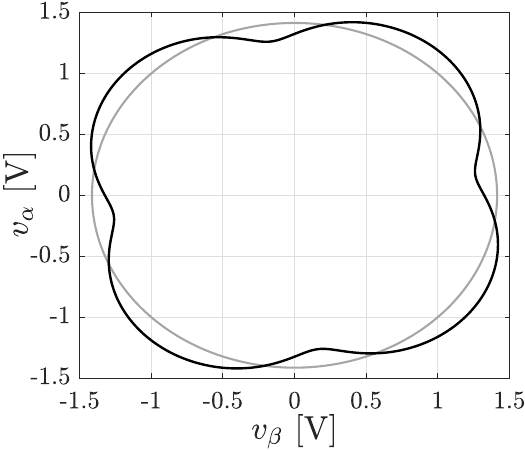}\label{fig:harmtrajex}}
  \caption{Comparison of voltage trajectories of (a) a sinusoidal unbalanced voltage signal (black line), and (b) a non-sinusoidal balanced voltage signal (black line) with respect of a sinusoidal balanced voltage signal (gray line).}
  \label{fig:illustration}
\end{figure}

In stationary conditions, i.e. $\partial_t\voltage=\bs{0}_3$, and no distortion, i.e. $\bs{R}=\bs{0}_{3,3}$, the instantaneous frequency is given by:
\begin{equation}
  \label{eq:stationary}
  \wu = \frac{1}{2}(\nabla\times\voltage) = \frac{1}{2}\vorticity \, .
\end{equation}
Reference \cite{Lagrange_FM} also shows that vorticity is constant in all stationary cases and it is proportional to the fundamental frequency.
In this scenario the voltage trajectory rotates constantly at the fundamental frequency forming a perfect circle (that is, a curve with constant curvature $\kappa$) every period (see gray lines in Fig.~\ref{fig:ellipse}).

For an unbalanced, non-sinusoidal or transient voltages, $\bs{\omega}_r$ and/or $\bs{\omega}_t$ are not zero and the trajectory is not a circle anymore (see Fig. \ref{fig:illustration}).

Regardless the presence of harmonics or unbalances, the vorticity always represents the actual inherent rotation of the system that does not depend on the operating conditions.  However, according to \eqref{eq:lagrange}, to calculate the vorticity, one needs to calculate $\nabla \times \voltage$.  This implies the knowledge of the dependence of the velocity (voltage) with respect to the position (fluxes), namely of the expression $\voltage(t, \flux)$.  In practice, this function is not available and cannot be determined based on the sole measurements of $\voltage$.  In order to overcome this problem, in the following section, we leverage a theorem developed by Cauchy, which relates the vorticity with the average rotation.  This theorem permits avoiding calculating $\nabla \times \voltage$.

\section{Definition of Quasi-Steady State Frequency}
\label{sec:cauchy}

\begin{figure}[htb]
  \centering
  \includegraphics[scale=0.35]{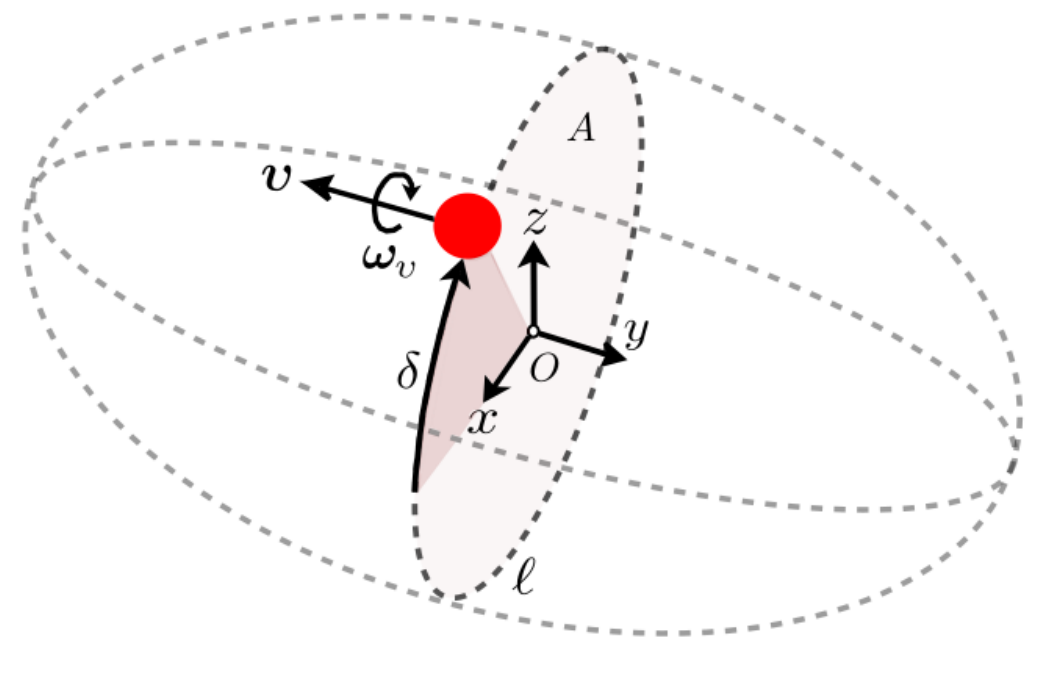}%
  \caption{Representation of the Cauchy's theorem that relates vorticity and the average of the angular speed, $\wu$, along the surface $A$, whose perimeter is the closed line $\ell$, for a point over a fluid line with velocity $\boldsymbol{\upsilon}$. Note that theorem applies for any arbitrary closed line.}
\label{fig:Cauchy}
\end{figure}

In \cite{Cauchy}, Cauchy developed a theorem that relates vorticity with average rotation, as follows:
\begin{equation}
  \label{eq:Cauchy} 
  \frac{1}{2\pi}\int_{0}^{2\pi} \wu (\delta) \, d\delta =
  \frac{1}{2}(\nabla\times\voltage)=\frac{1}{2}\vorticity \, ,
\end{equation}
where the left-hand term represents the average of the instantaneous rotation all over a closed line $\ell$ along a line of flow (see Fig.~\ref{fig:Cauchy}) and the right-hand term is half of the vorticity.  Note that \eqref{eq:stationary} represents a special case of \eqref{eq:Cauchy}, that is, the case for which $\wu$ is a constant vector and, hence its average along $\ell$ coincides with its punctual value.

We now have a relationship between $\wu$ and $\vorticity$ that does not involve $\bs{\omega}_r$ and $\bs{\omega}_t$.  However, we have shifted the problem to calculate $\vorticity$ to the problem of calculating the integral of $\wu$ with respect to the spatial angle $\delta$ along $\ell$.  The next step is to find a way to calculate the left-hand side of \eqref{eq:Cauchy} with the quantities that can be actually measured or estimated, which are all function of time.

With this aim,  we note that Cauchy's theorem is relevant for closed space curves that fulfill:
%
\begin{equation}
  \kappa_{\text{T}}=\oint_{\ell}^{}{\kappa}(s) \, ds=2\pi \,,
  \label{eq:kt_Jordan}
\end{equation}
where $\kappa_{\text{T}}$ is the total curvature, and $\kappa_{\text{T}}=2\pi$ is the necessary condition for a space curve to be a Jordan's curve, i.e., a simple closed curve \cite{Stoker}.
From arc length, curvature and geometric frequency definitions, in \eqref{eq:s}, \eqref{eq:k} and \eqref{eq:wu} respectively, \eqref{eq:kt_Jordan} can be rewritten as:
\begin{equation}
  \kappa_{\text{T}}=
  \oint_{T}^{}\kappa|\bs\upsilon|\,d\tau=\oint_{T}^{}
  \frac{|\voltage\times \voltage'|}{|\voltage|^{2}} \, d\tau=
  \oint_{T}^{}|\wu| \, d\tau  \, ,
  \label{eq:period_Jordan}
\end{equation}
and equaling \eqref{eq:kt_Jordan} with \eqref{eq:period_Jordan}, we obtain:
\begin{equation}
  \label{eq:2pi}
  \oint_{T}^{}|\wu| \, d\tau = 2\pi \, .
\end{equation}

Equation \eqref{eq:2pi} not only gives a condition for a space curve to be a Jordan's curve but it also provides a geometrical definition of the period, $T$, of a signal.  If the trajectory of a dynamical system generates a Jordan's curve, there exists an associated period defined as the time that it takes, for a given initial condition at time $t_0$,  to close the trajectory curve:
\begin{equation}
  \boxed{T=\text{inf}\,\left\{t: t>t_0\, , \int_{t_0}^{t+t_0}|\wu| \, d\tau= 2\pi \right\}} 
  \label{eq:Tdef}
\end{equation}

The information obtained with \eqref{eq:Tdef} is twofold.  On the one hand, if the curve is closed, one can define its equivalent period $T$ and the existence of the period implies the existence of an associated frequency.  On the other hand, if the trajectory is not closed, the period and thus the frequency cannot be defined.  Equation \eqref{eq:Tdef} is key for the developments of this work because: (i) it provides a way to obtain the period $T$ through the evaluation of the instantaneous frequency; and (ii) it can be calculated only with voltage measurements.

It is important to note that the mere existence of a period does not imply the applicability of Cauchy's theorem.  If the curve is a closed curve with double points, it has an associated period but it is beyond the Jordan's curves.  Hence, the Cauchy's theorem is not directly applicable and the associated period cannot be measured through the condition in (\ref{eq:period_Jordan}).  This issue is further discussed in Section \ref{sub:ex:harmonics}.

Once the period is evaluated with \eqref{eq:Tdef}, the average of the instantaneous rotation can be redefined in time-domain in the Cauchy's theorem, in the form of \eqref{eq:Cauchy}, leading to the proposed definition of \ac{qss} frequency, hereinafter indicated with $\wqss$:
\begin{equation}
  \label{eq:wqss}
  \boxed{\wqss = \frac{1}{T}\oint_{T}^{}{\wu} \,d\tau}
\end{equation}

The integral that appears in \eqref{eq:wqss} effectively operates averaging out the distortions ($\bs{\omega}_r$) and time-varying components ($\bs{\omega}_t$) that are included in the instantaneous frequency.  What remains after the calculation of \eqref{eq:wqss} is the sought fundamental frequency of the measured signal.  Note that $\wqss$ is time varying if the period $T$ is time varying. Hence, $T$ has to be updated for each evaluation of $\wqss$.   However, using a mobile window, the estimation of $\wqss$ can be obtained with the same rate as the sampling of the voltage measurements.  Finally, note that, in \eqref{eq:wqss}, $\wu$ is assumed to be calculated using \eqref{eq:wu}, which is defined for smooth voltages, except at most a finite number of discontinuities.   Moreover, the integral in \eqref{eq:wqss} has to be intended as a principal value integral.

\section{Time Derivative of Circulation as a Periodicity Condition}
\label{sec:circulation}

In the previous section, it has been assumed that the trajectory along which $\voltage$ is calculated forms a Jordan's curve.  In this section, we provide a metric, independent from the condition in (\ref{eq:Tdef}), to determine if this assumption is satisfied.
For this purpose, we utilize the concept of circulation along a smooth closed curve $\ell$ that was originally introduced for the study of fluid dynamics \cite{Chorin}.  Let $\Gamma$ be the circulation, then using once again the analogy between an electrical voltage and a velocity, one has:
\begin{equation}
  \Gamma = \oint_{\ell} \voltage \, ds \, .
\end{equation}
Then, applying Stokes' theorem, which relates the line integral of a vector to the surface integral of its curl \cite{Kambe}, the circulation can be written as:
\begin{equation}
  \Gamma = \oint_{\ell}\voltage \, ds =
  \int_{A}(\nabla\times\voltage) \, dA =
  \int_{A}\vorticity \, dA \,,
  \label{eq:circulation}
\end{equation}
where $A$ is an open surface bounded by $\ell$ and the right-hand term is defined as the flux of the vorticity \cite{kundu2003fluid}.

Next, we use Green's theorem, which states that, given two functions, ${p}(x,y)$ and ${q}(x,y)$, with continuous partial derivatives, the following identity holds \cite{riley2006mathematical}: 
\begin{equation}
  \oint_{\ell}(p dx+ q dy) =
  \int_{A}\left(\frac{\partial p}{\partial x}-
  \frac{\partial q}{\partial y}\right) dA \, .
  \label{eq:GREENS}
\end{equation}
The application of Stokes' and Green's theorems are here used to define the circulation as a linear integral with respect of the well known plane coordinates.  Green's theorem refers to a two dimensions, i.e., vorticity is defined in the normal direction with respect of the $(x,y)$ plane.  In three or more dimensions, the theorem can generally be applied to each combination of two coordinates separately.

In the electric domain, by choosing $p=\upsilon_\alpha(\varphi_\alpha,\varphi_\beta)$ and  $q=\upsilon_\beta(\varphi_\alpha,\varphi_\beta)$, the left-hand term of \eqref{eq:GREENS} is rewritten as:
\begin{equation}
  \begin{aligned}
    \oint_{\ell}(\upsilon_{\alpha}d\varphi_{\alpha}+\upsilon_{\beta}d\varphi_{\beta}) &=
    \oint_{T}(\upsilon_{\alpha}\upsilon_{\alpha}d\tau+\upsilon_{\beta}\upsilon_{\beta}d\tau) \\ &= \oint_{T}|\voltage|^{2}d\tau \, .
  \end{aligned}
  \label{eq:int_v2}
\end{equation}
Then, using \eqref{eq:GREENS}, given the left-hand term defined in \eqref{eq:int_v2}, in equation \eqref{eq:circulation}, the circulation is obtained as:
\begin{equation}
  \begin{aligned}
    \Gamma=\oint_{T}|\voltage|^{2}d\tau &= \int_{A}\left(\frac{\partial \upsilon_{\alpha}}{\partial \varphi_\alpha}-\frac{\partial \upsilon_{\beta}}{\partial \varphi_\beta}\right)dA \\
    &=\int_{A}\vorticity\cdot dA \, .
  \end{aligned}
  \label{eq:circulation_final_def}
\end{equation}
At this point, $\Gamma$ can be defined exclusively in terms of voltage.

We now use Kelvin's circulation theorem that states that the rate of change of circulation around a closed contour is zero \cite{kundu2003fluid}.  The relevance of this theorem for our work is that, given a measured period, it verifies whether the assumption that such a period exists is valid.
Applying this theorem to \eqref{eq:circulation_final_def}, the condition for a signal to represent a closed trajectory is:
\begin{equation}
  \Gamma' = \oint_{T}(|\voltage|^2)'\, d\tau = 0 \, .
  \label{eq:Kelvin}
\end{equation}

The time derivative of the circulation allows determining if the curve is periodic and, consequently, whether frequency can be defined.
Thus, we propose to use \eqref{eq:Kelvin} as a metric to evaluate the scope of validity of the \ac{qss} frequency. 

In practice, it is not possible to obtain exactly zero, even if the signal is in effect periodic, due to measurement and numerical errors.  We propose thus to use a  threshold, say $\epsilon$, and argue that the \ac{qss} frequency has the meaning of ``fundamental frequency'' if:
\begin{equation}
  \label{eq:metric}
  \boxed{\Gamma' = \oint_{T}(|\voltage|^2)'\, d\tau \in[-\epsilon,\epsilon]}
\end{equation}
While in principle arbitrary, the choice of $\epsilon$ is quite straightforward as $\Gamma'$ is very small if the curve is periodic and increases quickly as periodicity is lost.  We duly illustrate this point in the case study discussed in Section \ref{sec:casestudies}.

\section{Illustrative Examples}
\label{sec:examples}

To provide a clearer understanding of the presented concepts of \ac{qss} frequency and time-derivative of circulation, in this section, we present various analytical and numerical examples.

\subsection{Stationary balanced sinusoidal voltage}

Under stationary and balanced conditions, the voltage components of a three-phase system, in the \Clarke{} stationary frame, are given by:
\begin{equation*}
  \voltage=(\upsilon_\alpha,\upsilon_\beta,\upsilon_\gamma)^T=(V\cos{\theta},V\sin{\theta}, 0)^T\,,
\end{equation*}
being its magnitude:
\begin{equation*}
  |\voltage|=\sqrt{\voltage\cdot\voltage}=\sqrt{V^2\left(\cos^{2}{\theta}+\sin^{2}{\theta}\right)}=V \,,
\end{equation*}
and its time derivative:
\begin{equation*}
  \voltage'=\wo(-V\sin{\theta},V\cos{\theta},0)^T\,,
\end{equation*}
where $\theta=\wo t+\phi$ for a given initial shift $\phi$.

The instantaneous frequency, calculated with \eqref{eq:wu}, is:
\begin{equation*}
  \begin{aligned}
    \wu &=
    \frac{\wo}{V^2}\begin{pmatrix}
      V\cos{\theta}  \\
      V\sin{\theta}   \\
      0
    \end{pmatrix} \times \begin{pmatrix}
      -V\sin{\theta}   \\
      V\cos{\theta}   \\
      0
    \end{pmatrix} 
    =
    \wo \uvec{\gamma} \, .
  \end{aligned}
\end{equation*}
where $\uvec{\gamma} = [0, 0, 1]^T$.  From the Jordan's curve condition in \eqref{eq:period_Jordan}, the period of the signal can be calculated from the following identity:
\begin{equation*}
  \kappa_{\text{T}} =\oint_T|\wu|\,d\tau=\int^T_0\wo\,d\tau=\wo T=2\pi \,.
\end{equation*}

From which, according to the definition in \eqref{eq:Tdef}, the period will be equal to $T=2\pi/\wo$.
Once the period is obtained, the \ac{qss} frequency can be calculated as follows:
\begin{equation*}
  \begin{aligned}
    \wqss &= \frac{1}{T}\oint_T\wu \, d\tau \\
    &= \frac{\wo}{2\pi}\int^T_0\wo \uvec{\gamma} \, d\tau =
    \frac{\wo^2}{2\pi}T \uvec{\gamma} =
    \wo \uvec{\gamma} \,.
  \end{aligned}
\end{equation*}
As expected, if voltage is balanced, sinusoidal and stationary, the instantaneous frequency and the magnitude of the \ac{qss} frequency is equal to the fundamental frequency of the voltage, namely $\wo$.
Then, the time derivative of the circulation gives:
\begin{equation*}
  \Gamma' = |\voltage(T)|^2-|\voltage(0)|^2 = V^2-V^2 = 0 \, ,
\end{equation*}
which confirms that the voltage is periodic and $\wqss$ is well defined. 

\vspace{-3mm}

\subsection{DC voltage}

From a geometric perspective, the trajectory of a DC voltage is defined by a straight line, or equivalently, a line with zero total curvature.  Consequently, the associated period tends to infinity as, theoretically, it will take an infinite amount of time for the total curvature to reach the value of $2\pi$.  Thus, the \ac{qss} frequency and, equivalently, the local rigid body rotation are null.  This conclusion matches the results in \cite{Lagrange_FM} where DC voltage is described as an irrotational field.

\vspace{-3mm}

\subsection{Stationary unbalanced sinusoidal voltage}

Consider the following unbalanced voltage in \Clarke{} coordinates:
\begin{equation*}
  \voltage =
  (\upsilon_\alpha,\upsilon_\beta,\upsilon_{\gamma})^T =
  (V_\alpha\cos{\theta},V_\beta\sin{\theta},0)^T \, ,
\end{equation*}
with time derivative:
\begin{equation*}
  \voltage'=\wo(-V_\alpha\sin{\theta},V_\beta\cos{\theta},0)^T \, .
\end{equation*}

The instantaneous frequency measured with the geometric definition is:
\begin{equation*}
  \wu = \frac{\wo V_\alpha V_\beta}{|\voltage|^2} \uvec{\gamma}
  = \frac{\wo V_\alpha V_\beta}{V_\alpha^{2}\cos^{2}{\theta}+V_\beta^{2}\sin^{2}{\theta}} \uvec{\gamma} \, .
\end{equation*}

The period can be easily obtained with \eqref{eq:Tdef}.  However, for illustration, we use instead the calculation of the total arc length:
\begin{equation}
  \label{eq:st_ellipse}
  \begin{aligned}
    s_\text{T} &= s(T)-s(0) = \oint_{T} |\voltage|\,d\tau \\
    &=\int^{T}_{0} \sqrt{V_\alpha^{2}\cos^{2}{\theta}+V_\beta^{2}\sin^{2}{\theta}}\,d\tau \\
    &=\int^{T}_{0} \sqrt{V_\alpha^{2}(1-\sin^{2}{\theta})+V_\beta^{2}\sin^{2}{\theta}}\,d\tau \\
    &=V_\alpha\int^{T}_{0} \sqrt{1-\Big(1-\frac{V_\beta^{2}}{V_\alpha^{2}}\Big)\sin^{2}{\theta}}\, d\tau \\
    &=\frac{V_\alpha}{\wo}\int^{\wo T}_{0}  \sqrt{1-\Big(1-\frac{V_\beta^{2}}{V_\alpha^{2}}\Big)\sin^{2}{\theta}}\, d\theta \\
    &=\frac{V_\alpha}{\wo} \,\mathcal{E}\Big (\wo T\,|\,1-\frac{V_\beta^{2}}{V_\alpha^{2}} \Big ) \, ,
  \end{aligned}
\end{equation}
where $\mathcal{E}$ is used to represent the Legendre elliptic integral of the second kind of the variable $x$, given a parameter $k$, defined as $\mathcal{E}(x\,|\,k)=\int\left({1-k\sin^{2}{x}}\right)^{1/2}\,dx$ \cite{tableIntegral}.

It is well-known that the perimeter of an ellipse, which is the total arc length over the closed curve, is given by \cite{ellipse_perimeter}:
\begin{equation}
  s_\text{T}=a\mathcal{E}(2\pi|e^2)=a\mathcal{E} \Big (2\pi|1-\frac{b^2}{a^2} \Big )\, ,
  \label{general_st_ellipse}
\end{equation}
where $a$ and $b$ are the semimajor and semiminor axes, respectively, and $e$ is the eccentricity.

Comparison of \eqref{eq:st_ellipse} with \eqref{general_st_ellipse} proves that the trajectory of the voltages defines an ellipse parameterized by $a=V_\alpha/\wo$ and $b=V_\beta/\wo$.  Moreover, the comparison of the angle of both expressions shows that the period that ``closes'' the ellipse is $T=2\pi/\omega$.

Given the period, the \ac{qss} frequency is:
\begin{equation*}
  \begin{aligned}
    \wqss &=\frac{1}{T}\int^T_0|\wu| \uvec{\gamma} \,d\tau\\
    &= \frac{1}{T}\uvec{\gamma} \tan^{-1}\left(\frac{V_\beta\sin{(\wo T)}}{V_\alpha\cos{(\wo T)}}\right)\\
    &=\frac{1}{T}\uvec{\gamma} \tan^{-1}\left(\frac{V_\beta\sin{(2\pi)}}{V_\alpha\cos{(2\pi)}}\right)=
    \frac{2\pi}{T}\uvec{\gamma}=\wo\uvec{\gamma} \, .     
  \end{aligned}
\end{equation*}
This result shows that an unbalance in the voltage components causes the instantaneous frequency measurement to oscillate.  In contrast, the \ac{qss} frequency removes these oscillations and provides the actual fundamental frequency of the voltage.

The Jordan's curve condition is verified with \eqref{eq:Kelvin} as follows:
\begin{equation*}
   \begin{aligned}
    \Gamma' &=|\voltage(T)|^2-|\voltage(0)|^2 \\
    &=V^2_\alpha\left(\cos^{2}{(2\pi+\phi)}-\cos^{2}{(\phi)}\right) \\
   &+ V^2_\beta\left(\sin^{2}{(2\pi+\phi)}-\sin^{2}{(\phi)}\right)=0 \, .
   \end{aligned}
\end{equation*}
This identity states that any unbalanced sine wave voltage system describes a Jordan's curve every period $T$.  

\begin{figure}[htb]
  \centering
  \includegraphics[scale=0.875]{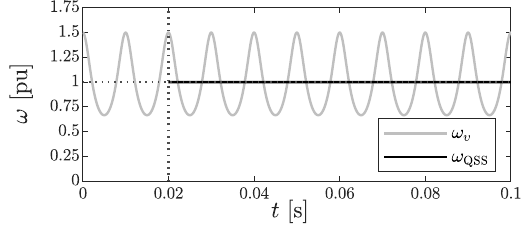}
  \vspace{0cm}
  \centering
  \includegraphics[scale=0.875]{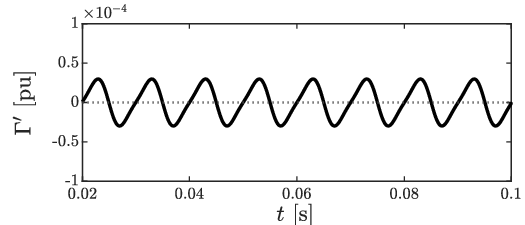}
  \caption{Stationary unbalanced sinusoidal AC voltage results for instantaneous frequency, \ac{qss} frequency and time derivative of circulation.}
  \label{fig:Unbalanced}
\end{figure}

For illustration, $|\wu|$, $|\wqss|$ and $\Gamma'$, for an amplitude unbalance of $V_\alpha=1.5V_\beta$, are shown in Fig.~\ref{fig:Unbalanced}.

The measured instantaneous frequency oscillates, forming a distorted sine shape.   The oscillations are visibly not symmetric with respect of the fundamental frequency.  However, the periods of the lower-amplitude oscillations are longer than the higher-amplitude oscillations, balancing the variations and resulting in an average value equal to the fundamental frequency.

For the same study case, in \cite{Lagrange_FM}, a different reference frame is used, leading to a different value of the vorticity. The intrinsic selection of reference frame of the \ac{qss} frequency measurement, ensures that the reference frame is such that the relative component of the frequency is zero. Note that the choice of a reference frame with this property is not unique.  However, despite the dependence of the vorticity on the reference frame, the approach proposed in this paper guarantees the reliability of the measured \ac{qss} frequency as it is measured from the geometric invariant, $\wu$, and the geometric period that is intrinsically related to this invariant.

This discussion is relevant to introduce the analysis of the most suitable reference frame to be used when measuring the \ac{qss} frequency, especially when relying on numerical rather than analytical calculations. By its definition in \eqref{eq:wqss} and the Jordan's curve condition in \eqref{eq:2pi}, the \ac{qss} frequency, measured from a static reference frame, is defined as follows:

\begin{equation*}
\begin{aligned}
  \wqss &= \frac{1}{T}\oint_{T}^{}{\wu} \,d\tau \\ &= 
  \frac{1}{T}\oint_{T}^{}{|\wu|\bs{e_{\wu}}} \,d\tau\\
  &=\frac{1}{T}\bs{e_{\wu}}\oint_{T}^{}{|\wu|} \,d\tau=\frac{2\pi}{T}\bs{e_{\wu}}=\wo\bs{e_{\wu}}
  \,,
\end{aligned}
\label{eq:wqss_static}
\end{equation*}
where $\bs{e_{\wu}}$ is the directional vector of $\wu$. The use of a static reference implies that: (i) the measurement of the \ac{qss} frequency can be reduced to the inverse of the measured period, reducing the use of numerical methods techniques and (ii) its value matches the fundamental frequency of the signal if exists. In fact, its value corresponds to the rigid-body rotation measured from a static reference frame. This idea is reinforced by the fact that, the \ac{qss} frequency definition matches with the fundamental coefficient of the Fourier's decomposition of the instantaneous frequency $\wu$.

\vspace{-3mm}

\subsection{Stationary balanced non-sinusoidal voltage}
\label{sub:ex:harmonics}

The case of the stationary balanced voltages described by a non-sinusoidal are here studied through the example of the addition of one harmonic of positive integer order $h$ and magnitude $V_h$. The voltages in the \Clarke{} frame are:
\begin{equation}
  \voltage =
  (\upsilon_\alpha,\upsilon_\beta,\upsilon_{\gamma})^T =
  \begin{pmatrix}
    V\cos{\theta}+V_h\cos{\theta_h}\\
    V\sin{\theta}+V_h\sin{\theta_h}\\
    0
  \end{pmatrix}
  \label{eq:harmonics}
  \,,
\end{equation}
where $\theta_h=h\wo t+\phi_h$ for a given initial shift $\phi_h$.
The time derivative of the voltage is:
\begin{equation*}
  \voltage' = \wo\left(-V\sin{\theta} -
  hV_h\sin{\theta_h},V\cos{\theta} +
  hV_h\cos{\theta_h},0\right)^T \,,
\end{equation*}
and the instantaneous frequency is:
\begin{equation*}
  \wu = \wo\frac{V^2+hV^2_h+(1+h)VV_h \cos{(\theta-\theta_h)}}{V^2+V^2_h+VV_h\cos{(\theta-\theta_h)}} \uvec{\gamma} \, .
\end{equation*}

We consider again the total arc length calculation.  The total arc length is:
\begin{equation*}
\begin{aligned}
    s_\text{T} &= \oint_{T} |\voltage|\,d\tau \\
    &=\int^{T}_{0} \sqrt{V^2+V^2_h+VV_h\cos{(\theta-\theta_h)}}\,d\tau\\
    &=-\frac{2\sqrt{V^2+V^2_h+VV_h}}{(h-1)\wo}\mathcal{E}\left( \frac{(1-h)}{2}\wo T    \,|\, \frac{2VV_h}{V^2+V^2_h+VV_h} \right)\\
    &=\frac{\sqrt{V^2+V^2_h+VV_h}}{\wo}\mathcal{E}\left( \wo T\,|\, \frac{2VV_h}{V^2+V^2_h+VV_h} \right)\\
    &=\frac{\sqrt{V^2+V^2_h+VV_h}}{\wo}\mathcal{E}\left( \wo T \,|\, 1-\frac{V^2+V^2_h-VV_h}{V^2+V^2_h+VV_h} \right) \,.
\end{aligned}
\end{equation*}
Then, the trajectory can be characterized as an equivalent ellipse parameterized by $a=(V^2+V_h^2+VV_h)^{1/2}\,/\wo$ and $b=(V^2+V_h^2-VV_h)^{1/2}\,/\wo$. According to the definition of the perimeter of an ellipse in (\ref{general_st_ellipse}), the period is proved to be $T=2\pi/\wo$.
Interestingly, while the parametrization of the ellipse is arbitrary, the results obtained with the original proposed formulation are not.

The time derivative of the circulation proves that the trajectory defines a Jordan's curve as follows:
\begin{equation*}
  \begin{aligned}
  \Gamma' &= |\voltage(T)|^2-|\voltage(0)|^2 = \\
  &= VV_h\left( \cos{\left(2\pi(h-1)+\phi-\phi_h\right)} -
  \cos{\left(\phi-\phi_h\right)} \right) = 0 \, ,
 \end{aligned}
\end{equation*}
as long as $h$ is an integer.

As the trajectory is proved to define a Jordan's curve, the identity in (\ref{eq:kt_Jordan}) holds and, hence, the \ac{qss} frequency can be simply calculated as:
\begin{equation*}
    \wqss=\frac{1}{T}\uvec{\gamma}\oint_T|\wu|\,d\tau=\frac{2\pi}{T}\uvec{\gamma}=\wo\uvec{\gamma} \,.
\end{equation*}
This result verifies the capacity of the \ac{qss} frequency on clearing the oscillating terms from the harmonics of the instantaneous frequency and provide the actual periodicity of the signal.

\begin{figure}[htb]
  \centering
  \includegraphics[scale=0.85]{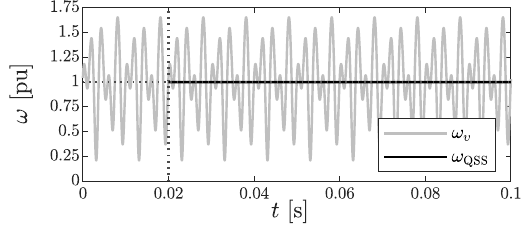}
  \centering
  \includegraphics[scale=0.85]{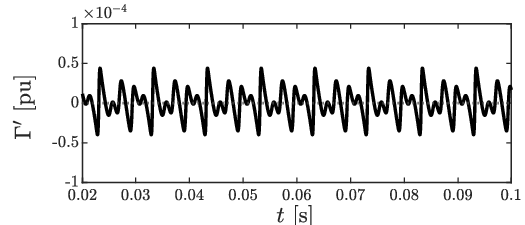}
  \caption{Stationary balanced non-sinusoidal AC voltage results for instantaneous frequency, \ac{qss} frequency and time derivative of circulation.}
  \label{fig:Harm}
\end{figure}

We illustrate the analytical results above with a numerical example.  The studied case considered harmonics orders $h=\{7,11\}$ with amplitudes of $V_h=\{5.83\%,3.71\%\}$ and phase angles of  $\phi_h=\{210^{\circ},330^{\circ}\}$. The results are displayed in Fig.~\ref{fig:Harm}.

\begin{figure}[htb]
  \centering
  \includegraphics[scale=0.85]{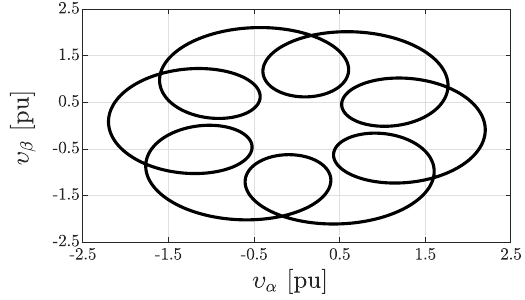}
  \caption{Stationary balanced non-sinusoidal AC voltage trajectory with the presence of high-level harmonics.}
  \label{fig:HarmHIGH}
\end{figure}

Non-sinusoidal voltages allow discussing the case of close curves that are not Jordan's curves.   If the harmonic level is high enough, in fact, the trajectory may present \textit{crunodes}, i.e. ordinary double points, before completing a period of the fundamental frequency.  This situation is shown in Fig.~\ref{fig:HarmHIGH} considering the harmonic $h=7$ with amplitude of $V_h=55.83\%$ and phase angles of  $\phi_h=210^{\circ}$.   

From a geometric perspective, harmonics represent a sequence of punctual deformations over the sinusoidal voltage trajectory defined by a circle if balanced or an ellipse if unbalanced.  In this scenario, the period, by definition, is not unique and the algorithm will measure the first one.  Consequently, the time derivative of circulation condition in (\ref{eq:Kelvin}) will not validate the physical existence of this measured period as a Jordan's curve period.

While the theoretical possibility of the existence of crunodes complicates the application of the approach proposed in this paper, we note that, from a practical point of view, the appearance of crunodes in the trajectory of a voltage measured in real-world power systems is highly unlikely.  For example, Standard IEEE 519 specifies that the acceptable individual voltage distortion for each harmonic is below the 5\% \cite{THD_STANDARD}.  In order to produce crunodes, for harmonics with this magnitude, the order of the harmonic has to be quite high, namely $h\ge20$.  The interested reader can find a geometrical justification of this condition in Appendix \ref{app:epi}.  In practice, harmonics of this order are filtered and their amplitude is well below 5\%.

\section{Case studies}
\label{sec:casestudies}

In this section, we discuss the estimation of the \ac{qss} frequency and the time derivative of the circulation as obtained during power system transients.  We also compare the \ac{qss} frequency with the instantaneous frequency estimation.  With this aim, we assess two cases: (i) a simulated EMT model with different operating scenarios; and (ii) measurements of a real-world voltage dip taken at a wind power plant installed in a Spanish distribution grid.

\subsection{EMT Simulation of the IEEE 39-Bus System}

The assessed model is the IEEE 39-bus system provided with DIgSILENT PowerFactory software tool.  Details on this systems and its set up can be found in in \cite{paradoxes}.  The results are measured at bus 26 and come from  EMT simulation, with a sample time of $10^{-5}$ s, under a secured three-phase fault in bus 4 from $0.2$ s to $0.3$ s.
In \cite{paradoxes}, it is concluded that the geometric frequency $\wu$ closely matches the instantaneous frequency  obtained with the \ac{pll}.  Thus, only the \ac{pll} frequency estimations and \ac{qss} frequency results are discussed here. 

\begin{figure}[t!]
  \centering
  \subfloat[]{\includegraphics[scale=0.49]{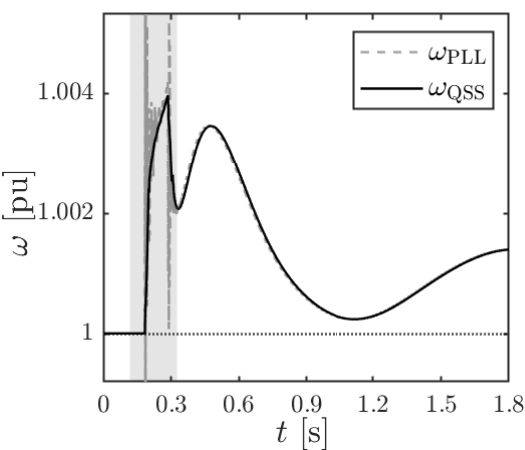}\label{IEEE_W_B}}
  \subfloat[]{\includegraphics[scale=0.49]{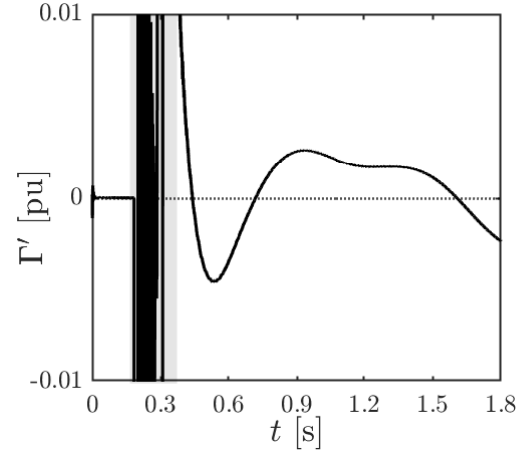}\label{IEEE_G_B}}\\
   \centering
  \subfloat[]{\includegraphics[scale=0.49]{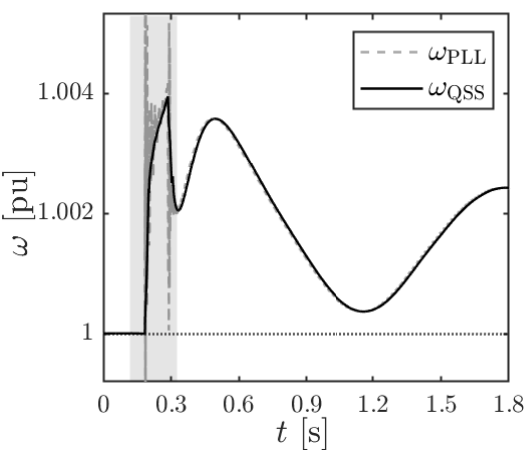}\label{IEEE_W_n}}
  \subfloat[]{\includegraphics[scale=0.49]{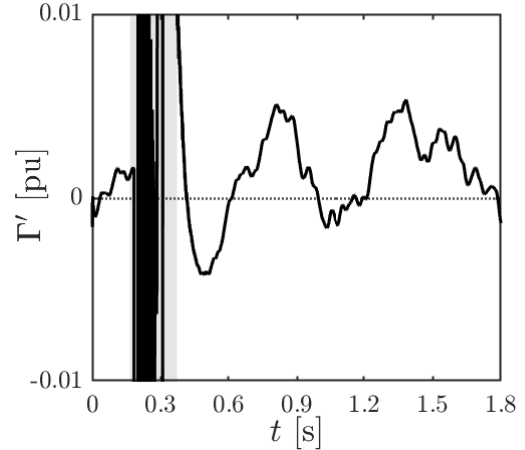}\label{IEEE_G_n}}\\
  \centering
  \subfloat[]{\includegraphics[scale=0.49]{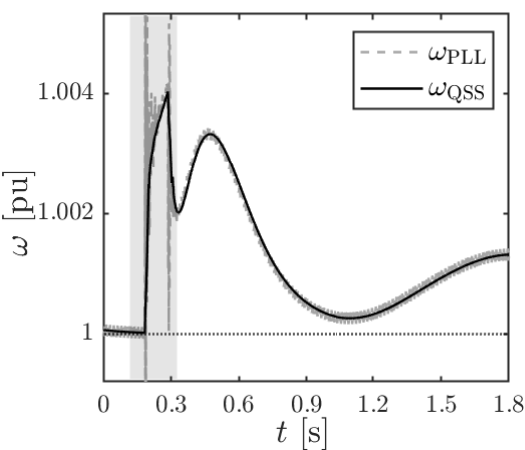}\label{IEEE_W_u}}
  \subfloat[]{\includegraphics[scale=0.49]{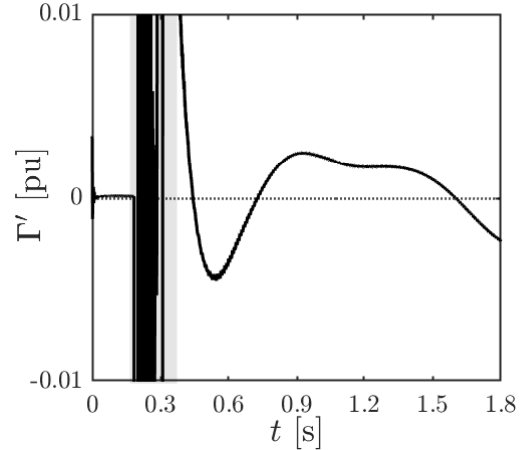}\label{IEEE_G_u}}\\
\centering
  \subfloat[]{\includegraphics[scale=0.49]{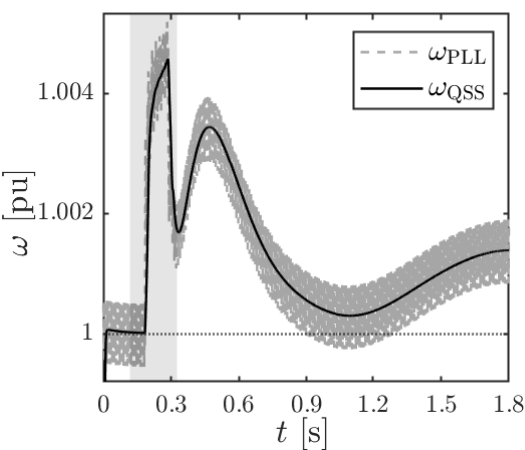}\label{IEEE_W_H}}
  \subfloat[]{\includegraphics[scale=0.49]{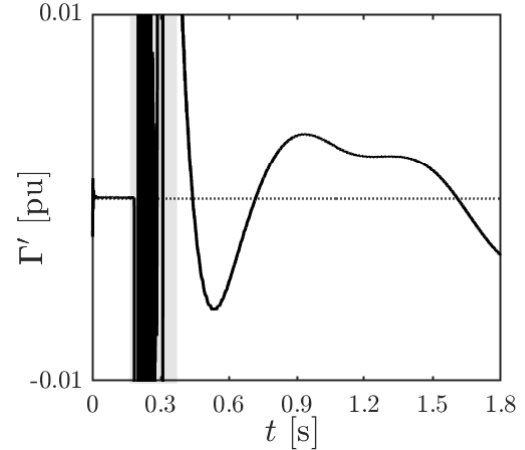}\label{IEEE_G_H}}
  \caption{Frequency and time derivative of the circulation of the voltage at bus 26 of the IEEE 39-bus model following a three-phase fault and its clearance after 100 ms: (a, b) balanced operation; (c, d) balanced operation with noise; (e, f) unbalanced operation; and (g, h) unbalanced harmonics current source.  Note that $\Gamma'$ shown in the right column refers to $\wqss$ and measures its scope of validity.  As it can be deduced from Section \ref{sec:circulation}, $\Gamma'$ cannot be calculated for the instantaneous frequency obtained with the \ac{pll}.}
  \label{fig:VSSK}
  \vspace{-3mm}
\end{figure}

Instantaneous frequency measurement ($\omega_{\text{\ac{pll}}}$), \ac{qss} frequency ($\wqss$) and time derivative of circulation ($\Gamma'$) are obtained for the following scenarios:
\begin{itemize}[leftmargin=4.5mm]
\item Balanced operation (Figs.~\ref{IEEE_W_B} and \ref{IEEE_G_B}).
\item Balanced operation with Gaussian noise in the voltage measurement (Figs.~\ref{IEEE_W_n} and \ref{IEEE_G_n}).
\item Unbalanced operation (Figs.~\ref{IEEE_W_u} and \ref{IEEE_G_u}).
\item Unbalanced $5^{\text{th}}$ and $7^{\text{th}}$ harmonics current source addition at bus 26 (Figs.~\ref{IEEE_W_H} and \ref{IEEE_G_H}).
\end{itemize}

Results indicate that the \ac{qss} frequency is consistent with the instantaneous estimation after the clearance of the fault, but it does not introduce a delay.  For balanced operation, no oscillations are produced in the instantaneous frequency measurement, and thus, both frequency results coincide before and after the occurrence of the fault.  For the scenarios with unbalanced and harmonics, on the other hand, the \ac{qss} frequency is capable of averaging out the oscillations of the instantaneous frequency and provide the fundamental frequency associated to the periodicity. 

The assessment of the time derivative of the circulation provides a deeper understanding of the nature of these results.  As mentioned in Section \ref{sec:circulation}, in order to evaluate this parameter, when working with discrete data, it is necessary to define a threshold that defines the interval within which the gap between the initial and the final point is small enough to be considered a closed curve.

By looking at the results of the balanced case under steady-state, we have set $\epsilon=10^{-2}$ pu as the threshold to decide if a curve is closed or not.  Figures \ref{IEEE_G_B}, \ref{IEEE_G_n}, \ref{IEEE_G_u} and \ref{IEEE_G_H} show that the results for all  scenarios lead to the same conclusion, as follows.  The measured values before and after the fault are below this threshold and, hence, prove that the trajectories can be defined by a Jordan's curve with a periodicity defined by the corresponding \ac{qss} frequency measurement.  On the other hand, during the fault -- grey-shaded area in the figures -- the time derivative of the circulation indicates that the trajectories, for the corresponding measured period, define an open curve. 

\begin{figure}[htb]
  \centering
  \subfloat[]{\includegraphics[scale=0.49]{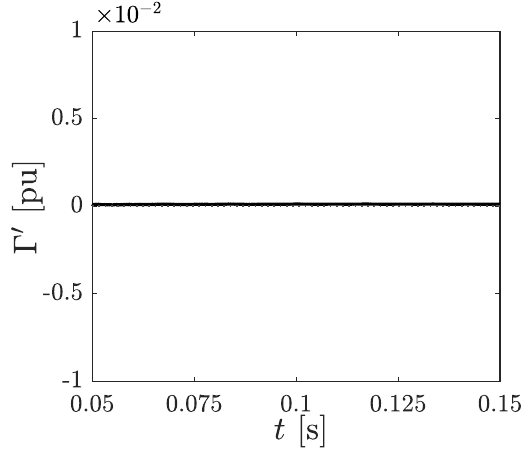}\label{detailedGAMMA}}
  \subfloat[]{\includegraphics[scale=0.49]{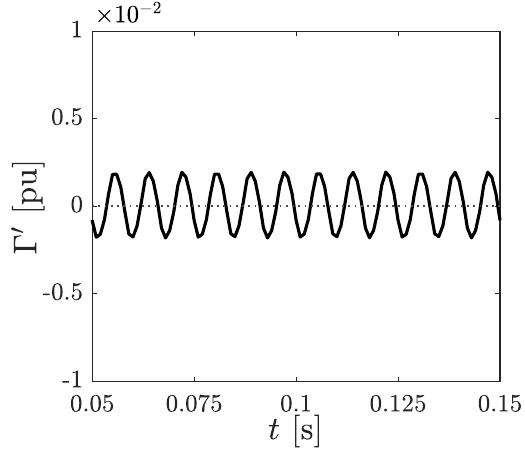}\label{detailedGAMMA_104}}
  \caption{Time derivative of circulation of the unbalanced scenario under steady-state operation from $0.05$ s to $0.15$ s for a sample time of (a) $10^{-5}$ s, and (b) $10^{-3}$ s.}
  \label{fig:numeric}
  \vspace{-3mm}
\end{figure}

Finally, numerical aspects of the evaluation of the time derivative of the circulation are discussed.  As it is to be expected, its accuracy relies on the sampling rate of the measurements and the numerical methods utilized to calculate derivatives and integrals, as well as to filter the noise.  However, based on several tests that  have been carried out, it is observed that $\Gamma'$ tends to change abruptly and of various orders of magnitude when the signal ceases to be periodic, whereas it tends to remain ``small'' when \ac{qss} conditions are satisfied.  For example, Fig.~\ref{fig:numeric} illustrates the effect of different sampling times on the evaluation of $\Gamma'$ in steady state conditions for the unbalanced case of the IEEE 39-bus systems discussed above.  Despite the sampling time changes of 2 orders of magnitude, $\Gamma'$ remains well below the threshold utilized to decide whether the signal is periodic.  

\subsection{Voltage Measurements at a DFIG Wind Generator Bus}
\label{sec:real}

To further evaluate the proposed quantities and their applicability to power systems, this section presents results based on real-world measurement data. Data were obtained at a sampling frequency of $10.5$ kHz from the stator of a \ac{dfig}, with nominal values of $690$V and $2$ MW, from a wind power plant in Moralejo, Spain.

\begin{figure}[htb]
  \centering
  \includegraphics[scale=0.85]{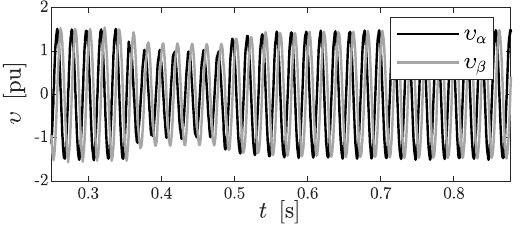}
  \caption{Voltage dip event measured at the stator of a \ac{dfig} of a wind power plant in Moralejo, Spain. Measurements have been reported in the \Clarke{} frame and normalized.}
  \label{fig:v_real}
\end{figure}

\begin{figure}[htb]
  \centering
  \subfloat[]{\includegraphics[scale=0.49]{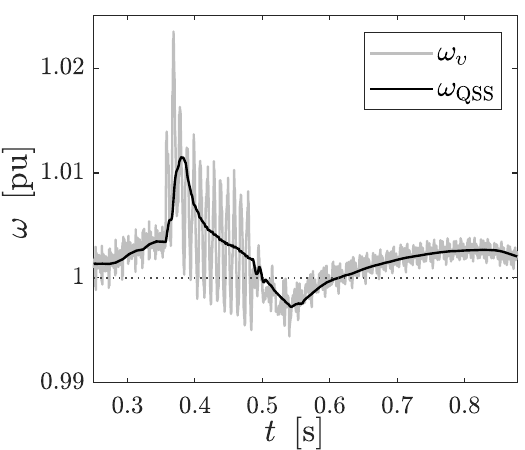}\label{real wv}}
  \subfloat[]{\includegraphics[scale=0.49]{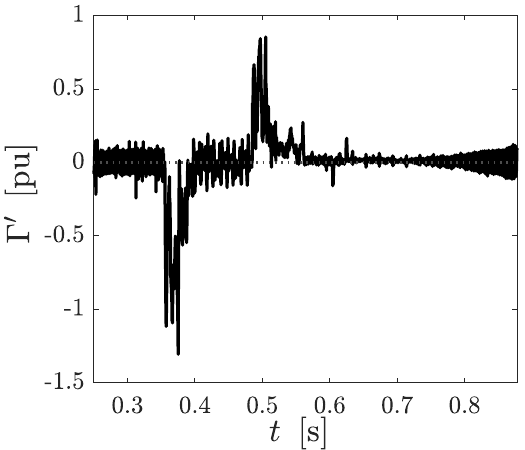}\label{real circ}}
  \caption{Instantaneous and \ac{qss} frequency (a) results of wind generator bus voltage measurements; (b) time derivative of the circulation, $\Gamma'$. }
  \label{fig:real wv and circ}
\end{figure}

The voltage measurements, which capture a voltage dip event, are shown Fig.~\ref{fig:v_real}, using the \Clarke{} frame.  In this scenario, the instantaneous frequency has been estimated with the geometrical formula in \eqref{eq:wu} (see \cite{milano2021geometrical}).  The result has been processed through a first-order Butterworth filter and plotted in Fig.~\ref{real wv} together with the \ac{qss} filter results.  Then, Fig.~\ref{real circ} shows the $\Gamma'$ metric.

As expected, the \ac{qss} frequency is able to remove the oscillations of $\wu$ and represents the slow-varying fundamental frequency associated to the periodicity of the signal.

\begin{figure}[htb]
  \centering
  \includegraphics[scale=0.85]{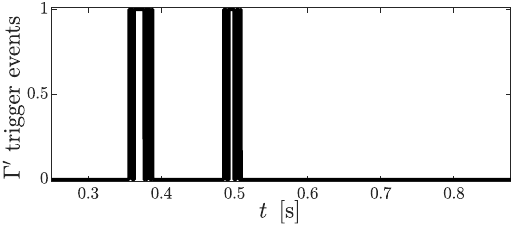}
  \caption{Boolean results of the condition \eqref{eq:metric} for $\epsilon=0.3$ pu for the real wind generator bus voltage measurements study case.  1 indicates that $|\Gamma'| > \epsilon$, 0 indicates $|\Gamma'| \le \epsilon$.}
  \label{fig:circ-TRIG}
\end{figure}

The evaluation of $\Gamma'$ for this case study allows us further discussing the election of a proper threshold $\epsilon$. With this aim, we evaluate the circulation in stationary conditions and define a proper threshold accordingly.  Due to noise in the measurements, in this case, we use  $\epsilon=0.3$ pu.  Figure \ref{fig:circ-TRIG} shows the instants when this threshold is exceeded (i.e., $\Gamma' > \epsilon$) at the beginning and the end of the voltage dip event.

Comparison between Figs. \ref{fig:circ-TRIG} and \ref{fig:v_real} indicate that \ac{qss} does not have a physical meaning during the transient periods after and before the voltage dip.  These results justify the importance of a proper threshold selection.  It is interesting to note that the abrupt variation of $\Gamma'$ during non-periodical conditions with respect to normal operating conditions facilitates the definition of the threshold.  This feature suggests that the loss of periodicity in the signal introduces structural changes in the shape of the circulation.  We aim at further investigating this point in future work.

\subsection{Voltage Measurements at a PV installation}
\label{sec:real_pv}

\begin{figure}[htb]
  \centering
  \includegraphics[scale=0.85]{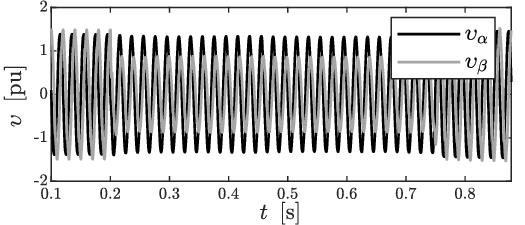}
  \caption{Unbalanced voltage dip event measured at PV installation in Moralejo, Spain. Measurements have been reported in the \Clarke{} frame and normalized.}
  \label{fig:v_pv}
\end{figure}

This subsection expands the real-world study cases by the assessment of a $10.5$kHz sampling rate voltage measurements of a PV installation in Moralejo, Spain.
Measurements, in the \Clarke{} frame are shown in Fig.~\ref{fig:v_pv}. The captured data includes an unbalanced operation under steady-state and a voltage dip between, approximately, $0.2$ and $0.8$ s on the $\beta$-axis.

\begin{figure}[htb]
  \centering
  \subfloat[]{\includegraphics[scale=0.49]{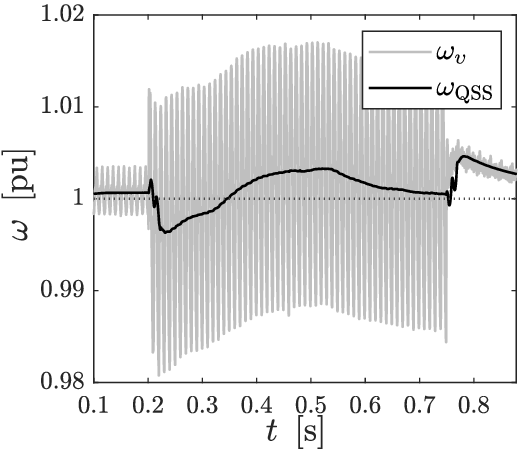}\label{real wpv}}
  \subfloat[]{\includegraphics[scale=0.49]{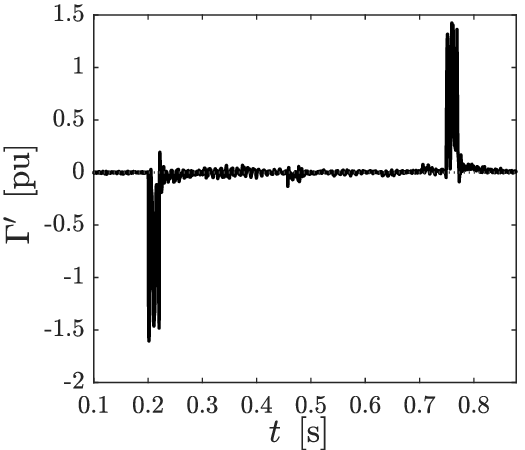}\label{real circpv}}
  \caption{ Instantaneous and \ac{qss} frequency (a) results of PV installation voltage measurements; (b) time derivative of the circulation, $\Gamma'$. }
  \label{fig:real pv wv and circ}
\end{figure}

Similarly to the previous scenario, instantaneous frequency has been estimated by means of \eqref{eq:wu} and the result has been filtered through a first-order Butterworth filter.  Figure \ref{real wpv} presents both instantaneous and \ac{qss} frequency results, while time derivative of circulation is plotted in Fig.~\ref{real circpv}.  As expected, the estimated \ac{qss} frequency is not affected by the oscillations due to unbalances and captures well the slow-varying fundamental value of the frequency.

\begin{figure}[htb]
  \centering
  \includegraphics[scale=0.85]{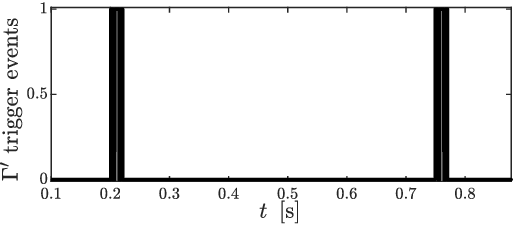}
  \caption{Boolean results of the condition \eqref{eq:metric} for $\epsilon=0.3$ pu for the real voltage measurements study case.  1 indicates that $|\Gamma'| > \epsilon$, 0 indicates $|\Gamma'| \le \epsilon$.}
  \label{fig:circ-TRIG-PV}
\end{figure}

From observing Figs.~\ref{real circ} and \ref{real circpv}, it appears that noise of the time derivative of circulation for this scenario is significantly lower.  However, we have used the same threshold $\epsilon=0.3$ pu in both scenarios for a fairer comparison.  Results of condition \eqref{eq:metric} are shown in Fig.~\ref{fig:circ-TRIG-PV} and show that the QSS frequency is not physically meaningful, as to be expected, in the periods when the voltage shows the most abrupt variations. 

It can be concluded that the clear step variations of $\Gamma^{\prime}$ makes quite permissive the choice of the threshold $\epsilon$ (as discussed in Section \ref{sec:real}).  In particular, the drastic jumps of $\Gamma^{\prime}$ allow choosing a relatively big threshold and help reduce the impact of measurement noise and of the numerical estimation  of $\wu$ required to calculate $\wqss$.

\section{Remarks}
\label{sec:remarks}

\subsection{On the \ac{qss} Frequency and Time Derivative of Circulation}

The following remarks on the properties of $\wu$ and $\Gamma'$ are relevant.
\begin{itemize}[leftmargin=4mm]
\item Analytical and numerical results validate the theoretical findings from section \ref{sec:cauchy} and \ref{sec:circulation}.  The \ac{qss} frequency measurement, based on this theoretical framework, is proved to be reliable on averaging out the instantaneous frequency oscillations.
\item The time derivative of circulation proves to be a consistent metric on verifying the periodicity of the signal and, consequently, the physical relevance of the frequency results.  In order to provide reliable information, for practical applications, it is necessary to define an appropriate threshold value with respect of the sample time of the measurements, as well as system conditions (e.g., harmonics and noise).
\item If periodicity exists, the \ac{qss} frequency measurement provides a value that accurately represents this periodicity through its fundamental period and frequency.
\item From the geometric definition of the period, an associated period does not exist for DC voltage and for extreme AC voltage transient periods (e.g., during a fault).  
\item With high harmonic distortion, the existence of the period might not be unique as the trajectory might represent multiple closed curves.
\end{itemize}


\subsection{On the Challenges of Frequency Estimation}
\label{sec:comparison}

Instantaneous frequency in power systems are typically estimated through \acp{pll} \cite{Monti2021}, while most \acp{pmu} measure the fundamental frequency via \ac{dft}-based algorithms \cite{milano2016advances}.  \acp{pll} are mostly utilized for the synchronization and the on-line control of power electronic converters, whereas \acp{pmu} are typically utilized for off-line power system monitoring and state estimation \cite{Monti2021}.

The proposed \ac{qss} frequency bridges these two approaches, with potential for both on- and off-line applications.  To support this statement, despite this paper is not focused on the design of the estimation of $\wu$ \textit{per se}, in the following, we discuss  potential intrinsic  challenges and limitations on the estimation of this quantity and how it compares with the frequency estimated with typical \acp{pll} and \acp{pmu}.

\begin{itemize}[leftmargin=4mm]

\item
For \ac{dft}-based algorithms, the sampling frequency is constrained by  the aliasing phenomenon \cite{romano2013interpolateddft}. In contrast, \ac{pll} and \ac{qss} estimations are limited only by the usual trade-off between accuracy and computational cost when discretizing a signal.

\item
Instantaneous frequency results might exhibit theoretical inconsistencies during certain operational states, e.g., oscillations under unbalanced conditions. This behavior is mitigated by filtering the output, which may, however, introduce undesirable delays.
Once the first period is measured, the \ac{qss} frequency averages out those oscillations by a mobile window approach. The length of this initial period can be shortened, though at the expense of accuracy. We plan to further investigate this trade-off when optimizing the \ac{qss} frequency estimation in future work (see Section \ref{sec:conclusions}). This mobile window approach is also used by \ac{dft} algorithms to estimate the fundamental frequency. 

\item
\ac{dft} algorithms are commonly based on static models that assume the waveform parameters to be constant throughout the assessed window. There also exist dynamic signal models approaches, e.g., Taylor–Fourier transform, but they imply a high computational complexity cost \cite{milano2016advances}. The computational complexity of \acp{pll} is defined by the control complexity whereas, for \ac{qss} frequency estimation, it relies on the instantaneous frequency measurement approach and the numerical methods. 

\item
The latency of \acp{pll} is influenced by the control loop parametritzation. Its design involves a trade off between latency and  accuracy of the measurements \cite{Paliwal2013}. This 
consideration also applies to \ac{qss} frequency estimation, given its reliance on instantaneous frequency measurement loop. Latency can be reduced in \ac{dft}-based algorithms by either lowering the sampling frequency, which might originate aliasing, or by shortening the window length, which potentially increases the spectral leakage effects \cite{milano2016advances}.
\end{itemize}

Finally, it is worth observing that there exist several ways to measure the instantaneous frequency, more or less sophisticated.  Advanced implementation of PLLs, for example, can return results better than the conventional synchronous-reference phase PLL that we have considered in the case study.  The objective of this work, however, is precisely to provide a theoretical framework on why certain methods return good results and on how to improve existing implementations.  Equally important, the paper also shows that the very definition of instantaneous frequency is theoretically inconsistent if certain geometrical properties (i.e., zero or almost zero circulation) are not satisfied.

\section{Conclusions}
\label{sec:conclusions}

This paper defines the novel concept of \ac{qss} frequency. 
This quantity fills the conceptual gap between stationary and instantaneous frequency. On the one hand, it does not require perfect stationary conditions and can be measured at any scenario.  On the other hand, it provides meaningful information regarding the nature of the signal's periodicity, if such periodicity exists.

To endow this frequency quantity with a formal analytical meaning, this paper also introduces the time derivative of circulation as a metric that determines the existence of periodicity and, consequently, of a fundamental frequency. This condition can be determined based solely on voltage measurements. 

The natural extension of this work is the testing of the \ac{qss} frequency and the time derivative of circulation metric in real-world and real-time applications and the comparison of the performance of electric systems with respect to other frequency measurements available in literature.  Potential applications include, among others, \ac{qss} frequency estimation for TSO control centers, microgrid controllers, EV fast-charging systems or its application for centralized or decentralized controls, and possibly, reinforced learning-based controllers.  In the case of the latter, \ac{qss} frequency can be used to train such controllers to improve their interpretation and anomaly detection, and integrated as an auxiliary safety signal in machine-learning-based frequency control architectures.  In this context, relevant applications are, for example, converter-based grid forming technologies, and detection of microgrid islanding and blackstart frequency anomalies.  Future research will also explore the optimization of the numerical methods to estimate the \ac{qss} frequency and, additionally, address the relation between the sample time of measurements and the necessary threshold to define this metric.  Finally, the potential of this metric as a technique to detect operational conditions, e.g., the presence of infra-harmonics  or multi-tone signals, will be further investigated.

\appendices

\section{Lagrange Derivative}
\label{app:lagrange}

In this appendix, part of the theory established in \cite{Lagrange_FM} about geometric frequency decomposition from Lagrange derivative equivalence is recalled to provide supplementary materials that support the here presented theory.

The Lagrange derivative for a given quantity, in this case $\voltage$, is defined as:
\begin{equation}
  \voltage'= \partial_t\voltage +
  \voltage \left( \voltage\cdot\nabla\right) \,.
  \label{eq:lagrange:der}
\end{equation}

From the triple cross product identity:
\begin{equation*}
  \bs{a}\times(\bs{b}\times\bs{c}) =
  \bs{b}(\bs{a}\cdot\bs{c})-\bs{c}(\bs{a}\cdot\bs{b}) \,,
\end{equation*}
the definition in (\ref{eq:lagrange:der}) can be rewritten, by identifying 
$\bs{a}=\bs{c}=\voltage$ and $\bs{b}=\nabla$, as:
\begin{equation*}
  \begin{aligned}
    \voltage' &= \partial_t\voltage+
    \frac{1}{2}\nabla(\voltage\cdot\voltage)-
    \voltage\times\left( \nabla\times\voltage\right) \\
    &= \partial_t\voltage+\frac{1}{2}\nabla|\voltage|^2+
    \left( \nabla\times\voltage\right)\times\voltage \,.
  \end{aligned}
\end{equation*}

The latter two terms can be rewritten by the Jacobian of the voltage, $\bs{J}=\voltage\nabla$, as follows:
\begin{equation*}
  \frac{1}{2}\nabla|\voltage|^2 =
  (\nabla\voltage)\voltage=\bs{J}^T \voltage \, ,
\end{equation*}
\begin{equation*}
  (\nabla\times\voltage)\times\voltage =
  (\bs{J}-\bs{J}^T)\voltage \,.
\end{equation*}

The Lagrange derivative, in terms of $\bs{J}$, is defined by:
\begin{equation*}
  \begin{aligned}
    \voltage' &=
    \partial_t\voltage+\bs{J}^T\voltage +
    (\bs{J}-\bs{J}^T)\voltage \\
    &=\partial_t\voltage+\bs{J}\voltage \, .
  \end{aligned}
\end{equation*}

The Jacobian of the voltage, as a square matrix, can be decomposed, from the Toeplitz transformation, into a symmetric matrix, $\bs{D}$, and a skew-symmetric matrix, $\bs{Q}$:
\begin{equation*}
  \bs{J} = \frac{1}{2}(\bs{J}+\bs{J}^T) +
  \frac{1}{2}(\bs{J}-\bs{J}^T) =
  \bs{D}+\bs{Q} \,.
\end{equation*}

Both matrices are tensors that represent different kind of dynamics. The symmetric matrix, $\bs{D}$, represents a pure strain motion, whereas the skew-symmetric matrix, $\bs{Q}$, represents the rigid-body angular rotation. 

The pure strain tensor can be further decomposed into two symmetric matrices by:
\begin{equation*}
  \bs{D} = \bs{S}+\bs{R} \,,
\end{equation*}
where $\bs{R}$ is the shear strain tensor and $\bs{S}$ is the normal strain tensor defined by the trace of $\bs{J}$:
\begin{equation*}
  \bs{S} =
  \frac{1}{n}\text{tr}(\bs{J})\bs{I}_n =
  \nabla\cdot\voltage \,,
\end{equation*}
where $n$ is the dimension of the voltage vector and $\bs{I}_n$ is the identity matrix of order $n$. The presented decomposition of $\bs{J}$ leads to the expression in (\ref{eq:lagrange}).

Finally, the Lagrange derivative can be rewritten, in terms of the presented tensors, as follows:
\begin{equation} 
  \label{eq:lagrange:final}
  \begin{aligned} 
    \voltage' &=\partial_t\voltage+\bs{S}\voltage+\bs{R}\voltage+\bs{Q}\voltage \\
    &=\partial_t\voltage+(\nabla\cdot\voltage)\voltage+\bs{R}\voltage+\frac{1}{2}(\nabla\times\voltage)\times\voltage \,.
  \end{aligned}
\end{equation}

In order to make the equivalence between the Lagrange derivative and the rotation component of the geometric frequency, the expression in (\ref{eq:lagrange:final}) is substituted in the definition in (\ref{eq:wu}).  Identifying vorticity definition, $\vorticity=\nabla\times\voltage$, rotation $\wu$ can be expressed by
%
\begin{equation*}
  \wu=\frac{\voltage\times \voltage'}{|\voltage|^2}=
  \frac{\voltage \times \partial_t\voltage}{|\voltage|^{2}}+
  \frac{\voltage \times (\bs{R}\voltage)}{|\voltage|^{2}}+
  \frac{1}{2}\frac{\voltage\times(\vorticity\times\voltage)}{|\voltage|^2}\,, 
\end{equation*}
where, by means of the triple cross product identity, and identifying $\bs{\omega_\tau}=(\vorticity\cdot\voltage)/|\voltage|$, where $\bs{\omega_\tau}$ is the torsional frequency (see \cite{wu2006vorticy}), the latter term can be rewritten as:
\begin{equation*}
\begin{aligned} 
   \frac{1}{2}\frac{\voltage\times(\vorticity\times\voltage)}{2|\voltage|^2}=&\frac{1}{2}\left(\frac{(\voltage\cdot\voltage)\vorticity}{|\voltage|^2}-\frac {(\vorticity\cdot\voltage)\voltage}{|\voltage|^2}\right) \\
   =&\frac{1}{2}\vorticity-\frac{1}{2}\bs{\omega}_\tau\frac{\voltage}{|\voltage|}\,,
   \end{aligned}
\end{equation*}
%
justifying the expression presented in \eqref{eq:lagrange}.


\section{Crunodes in Harmonic Voltage Trajectories}
\label{app:epi}

The trajectory of balanced voltages containing one harmonic can be geometrically constructed as an \textit{epitrochoid}.

An epitrochoid is a curve traced by a point, $P$, connected to a circle, with radius $r$, rolling on the exterior of a circumference of another circle with radius $R$ \cite{higher_plane_curves}.
Its parametric equations are defined as follows:
\begin{equation*}
  \begin{pmatrix}
    x\\y
  \end{pmatrix} =
  \begin{pmatrix}
    (R+r)\cos{\theta}-d\cos{(\dfrac{R+r}{r}\theta)}\\
    (R+r)\sin{\theta}-d\sin{(\dfrac{R+r}{r}\theta)}
  \end{pmatrix} \, ,
\end{equation*}
where $d$ is the distance between $P$ and the center of the circle with radius $r$.

Identifying $x=\upsilon_\alpha$ and $y=\upsilon_\beta$, the epitrochoid can be parametrized, from the voltage definition in \eqref{eq:harmonics}, as follows:
\begin{equation}
  \begin{aligned}
    d = V_h \, , \qquad
    r = \frac{V}{h} \, , \qquad
    R = V\left (1-\frac{1}{h} \right ) \, .
  \end{aligned}
  \label{eq:epiparam}
\end{equation}
Note that the sign of the right-hand term of the parametric equations is changed. This implies a shift on the direction of the rotation but does not change the traced trajectory.

By the assessment of the voltage time derivative, it is remarkable that $\upsilon^{\prime}_\alpha=\upsilon^{\prime}_\beta=0$ when $\theta=2n\pi/(h-1)$. This implies that, if $h$ is an integer, there are $n=(h-1)$ inner loops and critical points every period such that $\Delta\theta=2\pi$. 

\begin{figure}[htb]
  \centering
  \includegraphics[scale=0.49]{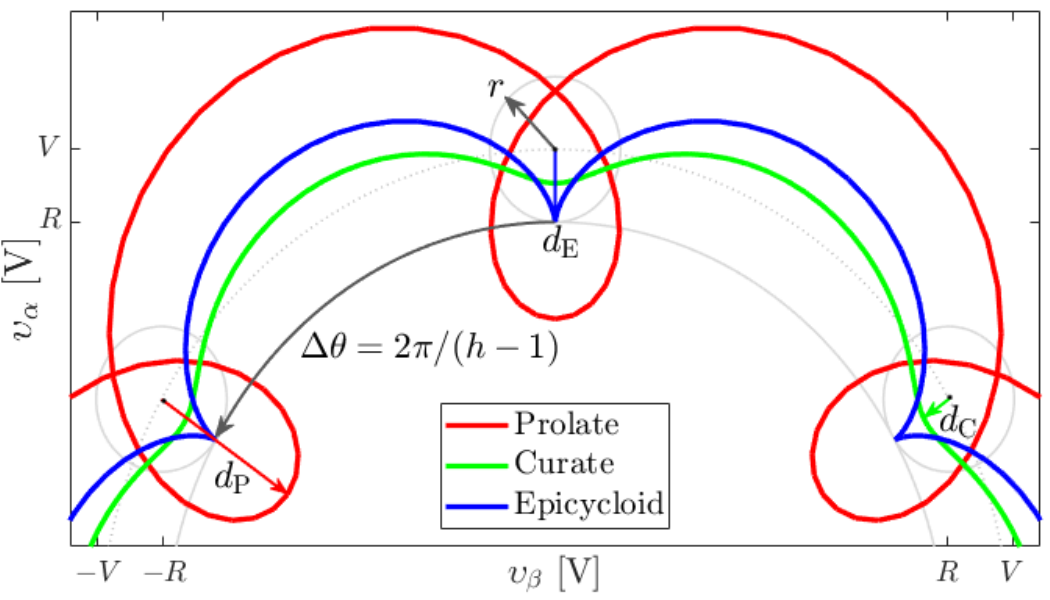}
  \label{epitrochoids}
  \caption{Detailed view of different trajectories of the voltage system in \eqref{eq:harmonics} for different values of $V_h$ such that trace different types of epitrochoids, $V_h=d_\mathrm{P}$ for a prolate, $V_h=d_\mathrm{C}$ for a curate, and $V_h=d_\mathrm{E}$ for an epicycloid.}
  \label{fig:epi}
\end{figure}

Epitrochoids can be classified in three different types, \textit{curtates} if $d<r$,  
 \textit{epicycloids} if $d=r$ and \textit{prolates} if $d>r$ \cite{types_epi}.
In Fig.~\ref{fig:epi} a detailed view of different voltage trajectories for different values of $d$ are shown.

For the critical points to be crunodes, the distance $d$ needs to be long enough to produce a loop, meaning the epitrochoid needs to be of the last type. 

Considering a set of balanced voltages with one harmonic, the prolate condition can be rewritten, using the parametrization in \eqref{eq:epiparam}, as:
\begin{equation*}
  \frac{V_h}{V}>\frac{1}{h} \, .
\end{equation*}
 Thus, for $V_h/V <= 1/h$, the period $T$ and, consequently, $\wqss$ are uniquely defined.   For example, given the edge case of $V_h/V=5\%$ specified in the standard \cite{THD_STANDARD}, this condition proves that the harmonic order needs to be $h\geq20$ to form crunodes.


\vfill

\end{document}